\documentclass[showpacs,preprintnumbers,amsmath,amssymb]{revtex4}
\usepackage{amsmath}
\usepackage{empheq}
\usepackage{graphicx}
\begin{document}

\newcommand{\be}{\begin{equation}}
\newcommand{\ee}{\end{equation}}

\title{The unified history of the viscous accelerating universe and phase transitions}

\author{A.V. Astashenok$^{1}$\footnote{E-mail: aastashenok@kantiana.ru},
S. D. Odintsov$^{2,3,4}$\footnote{E-mail: odintsov@ieec.uab.es},
A. S. Tepliakov$^{1}${\footnote{E-mail: ateplyakov@kantiana.ru}
}}

\medskip

\affiliation{$^{1}$Institute of Physics, Mathematics and IT, Immanuel Kant Baltic Federal University, 236041 Kaliningrad, Russia\\
$^2$Consejo Superior de Investigaciones Cient\'{\i}ficas, ICE/CSIC-IEEC,
Campus UAB, Carrer de Can Magrans s/n, 08193 Bellaterra (Barcelona) Spain\\
$^3$International Laboratory for Theoretical Cosmology, Tomsk State University
of Control Systems and Radioelectronics (TUSUR), 634050 Tomsk, Russia\\
$^{4}$ICREA, Passeig Luis Companys, 23, 08010 Barcelona, Spain}

\begin{abstract}
We propose the unified description of the early acceleration (cosmological inflation)
and the present epoch of so called ``dark energy''. The inflation can be described by cosmic fluid with van der Waals equation of state and with viscosity term. Viscosity leads to slow-roll inflation with the parameters such as the spectral index, and the tensor-to-scalar ratio
in concordance with observational data. Our next step is to modify this equation of state (EoS) to describe the present accelerated expansion. One can add the term into EoS so that the contribution of which is small for inflation but crucial for late-time acceleration. The key point of the model is possible phase transition which leads to decrease of the viscosity. We show that proposed model describes observational data about standard ``candles'' and correct dependence of Hubble parameter from redshift. Moreover, we propose the possible scenario to resolve dark matter problem.   
\end{abstract}

  \maketitle
\section{Introduction}

The observational data from Planck \cite{ade13,Akrami:2018odb} confirmed the inflation theory in its simple form \cite{linde08, linde14}. The non-Gaussian perturbations of some type are negligible and therefore simplest single-field models are valid. The results of Planck ruled out some alternative scenarios.

The inflation is very rapid expansion of the universe in an unstable state at the top of the effective potential. Scalar field slowly rolls down to
the minimum of effective potential.  Due to the expansion the universe became flat and very big. During the inflation the density perturbations are generated. These perturbations are inversely proportional to the velocity of scalar field decreasing $\dot{\phi}$. In slow-roll regime perturbations generated during the inflation have a spectrum very close to flat one. 

Although inflation usually is described in terms of scalar field theory one can use a perfect fluid model or modified gravity \cite{bamba14,nojiri07a,Brevik:2017juz}. In \cite{brevik17} the authors obtained conditions for preventing the occurrence of self-reproduction in inflation epoch. The case of multiple coupled viscous fluid was considered in  \cite{wang14,balakin11,nunes11}. The possibility of coupling between energy and dark matter was investigated in 
\cite{bolotin15}, and some examples of inhomogeneous viscous coupled fluids
were considered in \cite{bamba12,elizalde14,brevik15}. One can note also cosmological models with bouncing, which are
caused by an inhomogeneous viscous fluid
\cite{brevik14}. 

A number of papers is devoted to viscous fluid with so callled van der Waals EoS
\cite{capozziello02,kremer03,capozziello03,kremer04,khurshudyan14}. The van der Waals fluid model can account both for the early and late-time accelerated expansion stages of the universe. That is, inflation may be described by a van
der Waals fluid, with the specific properties of a cosmic fluid obeying the van der Waals equation of state having been analyzed \cite{vardiasli17}. Various
versions of the van der Waals equation can be considered in seeking a better match with the observational data of the Planck satellite \cite{jantsch16}. 

Another interesting cosmological phenomenon is accelerated expansion of the Universe \cite{Riess,Perlmutter}. From a theoretical viewpoint the cosmological acceleration can be caused by some fluid with
negative pressure and/or negative entropy (for review, see
\cite{bamba12,Dark-6,Cai:2009zp}). The nature of this fluid is unclear and late-time
accelerated expansion is dubbed also as the dark energy era. The
current observations coming from Planck \cite{Akrami:2018odb},
indicate that dark energy consist of nearly 70\% of total energy
density of the Universe \cite{Kowalski}. The
EoS parameter for the dark fluid, namely $w_{d}$, is negative:
 \be
w_\mathrm{d}=p_{d}/\rho_{d}<0, \ee where $\rho_{d}$ and $p_{d}$
are the dark energy density and pressure, correspondingly. However,
it is still not clear what is the precise value of $w_d$
\cite{PDP,Amman}, although the latest Planck data constrains
significantly the values that the EoS parameter can take.

The interesting question is whether it is possible the description not only of the early-time inflation but also of the late cosmological acceleration in unified way with using the viscous fluid. Note that viscous dark energy models as well as viscous inflationary models are considered in many papers  \cite{brevik94,brevik04,cataldo05,brevik06,brevik02,li09,brevik10a,sebastiani10,velten12,velten13,velten13a,bamba16,capozziello06a,nojiri07,bre,Odintsov:2018obx,Capozziello:2018mds}. 

We studied the possible unified description of early and late-time accelerated expansion of universe in terms of the van der Waals equation of state for cosmic fluid including viscous term. Slow-roll inflation is caused by viscous term in the equation of state being proportional to square of Hubble parameter i.e. the energy density if we neglect the contribution of matter and radiation. The exit from the rapid expansion is provided by decrease of viscosity in the form of a phase transition. One can add other term to the EoS of van der Waals fluid which contribution is negligible for inflation epoch but crucial for late-time acceleration. 

We calculate parameters of inflation for considered model and especially consider such values of parameters which are compatible with Planck data. One should stress the role of viscosity in considered model. The agreement with Planck data can be achieved only if viscosity term is present.

We start in the Section II from the basic cosmological equations and van der Waals equation for viscous fluid which plays the role of dark energy in our model. One assumes that viscosity depends from the energy density. Then we consider the possible unification of inflation with the era of matter domination. Section IV is devoted to the comparison of results of considered model with Planck data. We calculated spectral index and slow-roll parameters for the model. Finally the adopted EoS for the description of late time acceleration is considered. One can include in the equation of state such terms which don't affect the inflation parameters but lead to cosmological acceleration. The analysis shows that proposed model is compatible with observational data such as data about dependence ``magnitude-redshift''  for SN Ia and dependence of Hubble parameter from redshift. Some outlook is given in the conclusion.   

\section{Basic viscous universe}

We consider spatially flat
Friedmann-Lema\^{\i}tre-Robertson-Walker spacetime with the metric:
\begin{equation}
    ds^2=dt^2-a^2(t)(dx^2+dy^2+dz^2)
\end{equation}
In the natural system of units ($c=8\pi G=1$) the first Friedmann equation is
\begin{equation}
H^2=\frac{\rho}{3}, \label{1}
\end{equation}
where $\rho$ is the energy density, $H(t)=\dot{a}(t)/a(t)$ is the Hubble
parameter, $a(t)$ the scale factor.

The energy density is 
\begin{equation}
    \rho=\rho_d + \rho_m + \rho_r,
\end{equation}
where $\rho_m$ and $\rho_d$ are the density of matter and radiation correspondingly and $\rho_d$ is energy density of cosmic dark fluid. 

We take the following nonlinear inhomogeneous equation of state for this fluid:
\begin{equation}
p_d = w(\rho_d,t)\rho_d - H\zeta(H,t) + f(\rho_d). \label{3}
\end{equation}
Here $w(\rho,t)$ is thermodynamical parameter which depends from the energy
density and time. The second term describes bulk viscosity. The viscosity $\zeta(H,t)$ in general case depends from  Hubble
parameter and time (see \cite{bre}). We add also arbitrary function of density into equation of state $f(\rho)$. In the simplest case $f(\rho)=0$. Similar EoS is typical in theories of modified gravity \cite{capo}.

One can note that equation of continuity for cosmic fluid with equation of state (\ref{3}) is the same for non-viscous fluid:
\begin{equation}
\dot{\rho_d} +3H(\rho_d+p_d)=0. \label{4}
\end{equation}
Usually, in viscous cosmology, this equation would include a
term on the right hand side containing the bulk viscosity. But, we should
emphasize that the properties coming from viscosity are here included through the
inhomogeneous equation of state, instead of via a more standard bulk viscosity term.

We consider the van der Waals fluid model
\cite{vardiasli17}. As well known the van der Waals theory in hydrodynamics describes the gas when density is sufficiently large and due to this gas cannot be considered as the ideal one. In this case one should take into account the finite volume of molecules of the gas and viscosity.   

At first glance this situation is very far from the cosmological fluids. However one can assume that at the very early universe the phase transition takes place and viscosity term plays significant role in the inflation. This gives the ground to think of van der Waals fluid models as a serious and reasonable alternative in cosmological applications. Indeed, the corresponding equation fits very naturally in, as a possibility for an inhomogeneous equation
of state.

Following this approach we consider
thermodynamic parameter to be of the form
\begin{equation}
w(\rho_d,t)= \frac{w_{0}}{1-\beta \rho_d/\rho_c}.
\end{equation}
Here $w_0$ is the constant value. For $\rho_d<<\rho_c/\beta$ we have simply $w\approx w_{0}$. Parameter $\beta$ plays the role a critical thermodynamic parameter. The critical  value $\rho_c$ indicates that cosmic fluids change phases under
certain thermodynamic conditions. Whereas perfect fluids do not permit phase
transitions to occur, here the phase transition phenomenon can be occur
by means of the two-phase fluid described by the van der Waals equation.
For bulk viscosity the following dependence can be chosen:
$$
\zeta(H,t)=\xi(H)(H)^n.  \label{5}
$$
We assume that function $\xi(H)$ weakly depends from the Hubble parameter in the sense that $\frac{d\xi}{dH}<<n(H)^{n-1}$.

We consider the simplest case $n=1$ for the bulk viscosity. In this case equation of state (\ref{3}) takes the following form
\begin{equation}
p_d=\frac{w_{0} \rho_d}{1-\beta \rho_d/\rho_c} + f(\rho_d)-\xi \rho/3 \label{6}
\end{equation}
For inflationary acceleration epoch one can neglect the densities of radiation and matter and assume that $\rho\approx\rho_d$.
Next, using equations (\ref{4}) and (\ref{6}) one obtains the following
differential equation relating the scale factor to the energy density:
\begin{equation} 
a\left( 1-\beta \frac{\rho_d}{\rho_c}\right) \frac{d\rho_d}{da}+3\rho_d \left[ (1-\xi/3)(1-\beta\rho_d/\rho_c)+w_0 + \frac{f(\rho_d)}{\rho_d}(1-\beta\rho_d/\rho_c)\right]=0. \label{7}
\end{equation}

Let's assume that viscosity depends from the fluid density in the following way:
\begin{equation}\label{drop}
    \xi(\rho_d)=\xi_{0}-\Delta\xi \exp(-(\rho_d-\rho_0)^2/\Delta\rho^2),\quad \Delta\xi>0.
\end{equation}
If $\rho_d>>\rho_0, \Delta\rho$ we have $\xi\approx\xi_0$. In vicinity of $\rho_0$ viscosity drops down. {The main motivation for Eq. (\ref{drop}) is describing of the possible phase transition for viscous fluid when the value of viscosity sharply decreases.}  

For $f(\rho_d)$ we can choose for simplicity:
\begin{equation}
    f(\rho_d)=-\frac{\alpha\rho^2_d}{\rho_c}
\end{equation}

The equation (\ref{7}) can be rewritten in the following form:
\begin{equation}\label{eqln}
\frac{d\ln a}{dx}=-\frac{(1-\beta x)}{x(a_{1}x^2+a_{2}x+a_{3}+d(x))}.
\end{equation}
Here $x=\rho_d/\rho_c$ is dimensionless density. The coefficients $a_{i}$ are
$a_{1}=\alpha \beta, \, a_{2}=\beta \xi_0/3-\alpha -\beta, \, a_{3}=w_0-\xi_0/3 +1$. The function $d(x)$ is
$$
d(x)=\frac{1}{3}\Delta \xi \exp(-(x-x_f)^2/{\Delta x}^2)(1-\beta x), \quad \Delta x = \Delta\rho/\rho_{c}.
$$
For $x>>x_{f}$ the function $d(x)\approx0$ with good accuracy. For the epoch of cosmological expansion $x$ decreases and scale factor $a$ increases. Therefore r.h.s. of Eq.(\ref{eqln}) should be positive. 

Initially consider the case when $\Delta\xi=0$. 
If equation
$$
a_{1}x^2+a_{2}x+a_{3}=0
$$
has at least positive root $x_1$ and initial value of $x(0)>x_1$ the scale factor $a\rightarrow\infty$ for $x\rightarrow x_1$. Therefore we have the expansion according to the exponential law at large times with some constant energy density and $p_d/\rho_d\rightarrow -1$ for $t\rightarrow\infty$.

Possible sharp decrease of viscosity (one can consider this process as \textit{phase transition}) leads to decrease of the cosmological expansion rate and for some parameters to exit from the inflation. 

There are two possible scenarios of this phase transition:

1) $\rho_f\neq 0$ i.e. viscosity decreases in the range $\rho_f<\rho_d<\rho_c$ and then viscosity increases.

2) $\rho_f = 0$, viscosity decreases with density. 

For the illustration we consider the model with fixed parameters $\alpha=0.005$, $\beta=0.05$ and assume that $\rho(0)=\rho_c$, $\rho_0=0$. On Fig. 1 we give the dependence of e-foldings number $\ln(a/a_{inf})$ where $a_{inf}$ is initial scale factor. Due to the decrease of the viscosity inflation ends. The time of exit from the phase of fast expansion (exit on plateau) depends from $\Delta\xi$ and $\Delta x^2$. Hubble parameter sharply decreases (see Fig. 2). For illustration we choose such values of $\Delta\xi$ and $\Delta x^2$ that numbers of e-foldings lie withing realistic limits.      

\begin{figure}
    \centering
    \includegraphics[scale=0.4]{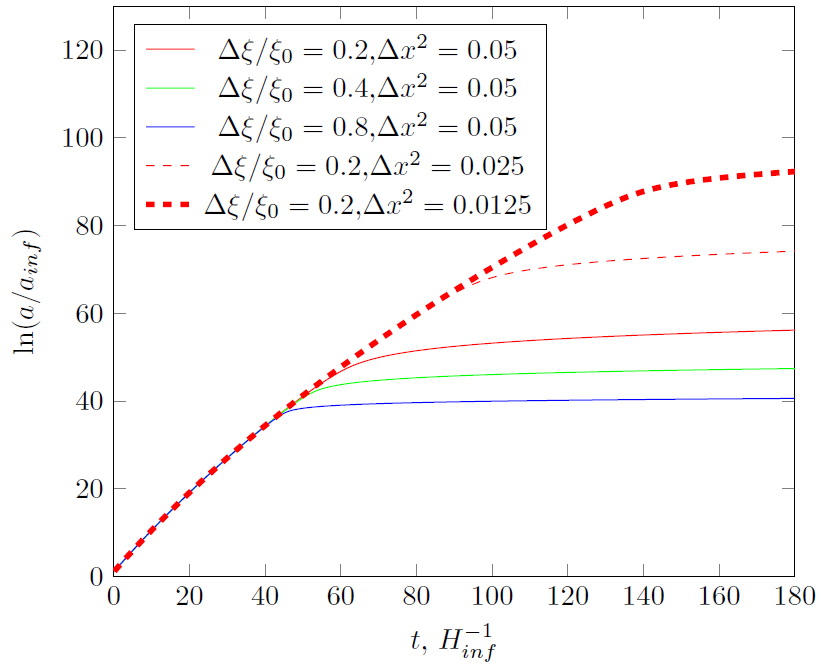}
    \caption{Dependence of e-foldings $\ln(a/a_{inf})$ from time (hereafter in units of $H_{inf}^{-1}$, $H_{inf}$ means value of Hubble parameter in the beginning of inflation) for various parameters of the model with equation of state (\ref{6}). Parameters $\alpha=0.005$, $\beta=0.05$ are fixed.}
    \label{fig:01}
\end{figure}

\begin{figure}
    \centering
    \includegraphics[scale=0.4]{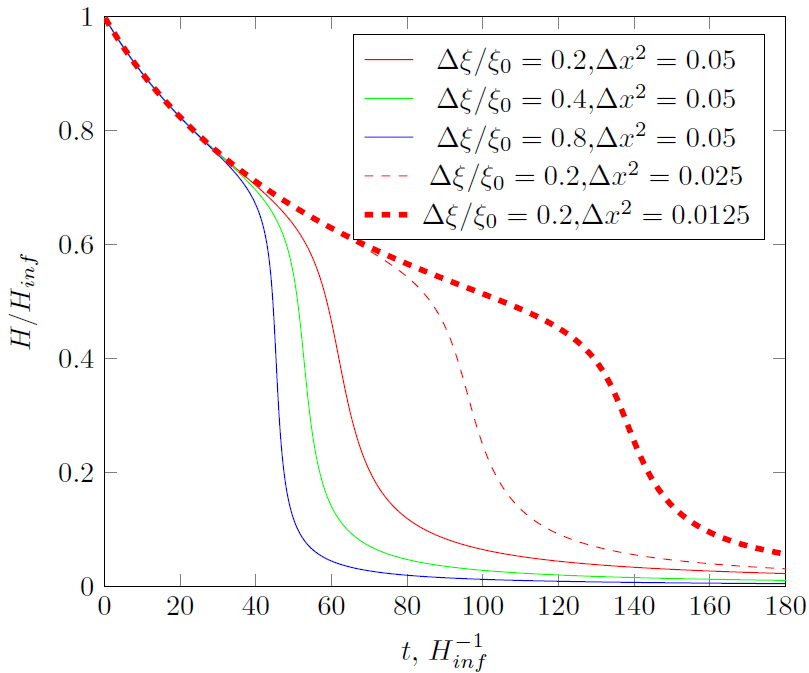}
    \caption{Dependence of Hubble parameter (in units of $H_{inf}$) from time for various parameters of the model with equation of state (\ref{6}).}
    \label{fig:02}
\end{figure}

\section{The unification of the inflation with matter domination epoch}

One can construct models in which dark fluid mimics dark matter with the effective value of state parameter close to $w\approx 0$. The asymptotic value of $w$ after inflation depends from parameter $\Delta \xi$ (for given $w_0$ and $\alpha$ and $\beta$. On Fig. 3 we depicted the dependence of effective equation of state parameter
$$
w=p_d/\rho_d
$$
from time for various values of $\Delta x^2$. For $\alpha=0.005$, $\beta=0.05$, $\xi_0=3.42$ we have asymptotic value $w\approx 0$ on exit from the inflation if $\Delta\xi=0.87\xi_{0}$. Phase transition leads to sharp increase of $w$ from the value nearly $-1$ to $0$. The change of $\Delta\xi$ leads to various asymptotical value of EOS parameter. Therefore our example can describe transition to matter domination epoch at least qualitatively.   

\begin{figure}
    \centering
    \includegraphics[scale=0.4]{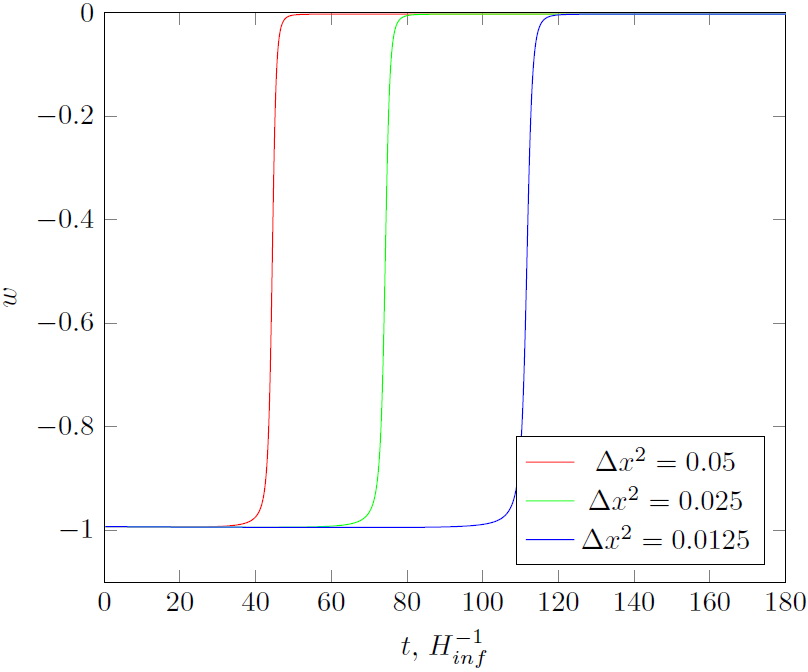}\\
    \includegraphics[scale=0.4]{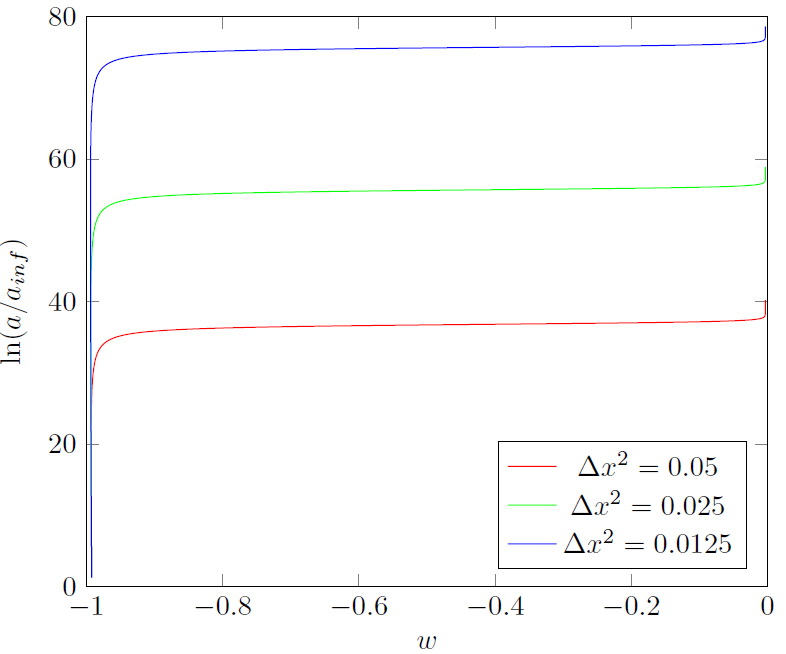}\\
    \caption{The dependence of the equation-of-state parameter $w=p_d/\rho_d$ from time (upper panel) and number of e-foldings as the function of $w$ (down panel). Other parameters of the model are $\alpha=0.005$, $\beta=0.05$, $w_0=0.145$, $\xi_0=3.42$, $\Delta\xi=0.87\xi_0$.}
    \label{fig:03}
\end{figure}

The possible interaction between viscous fluid and baryon matter also leads to interesting consequences. Let's consider the Universe which is filled with dark energy and baryon matter
$$
\rho=\rho_d + \rho_m. 
$$
and assume the interaction between matter and dark energy. Then the following equations for dark energy and matter density are satisfied
\begin{equation}
    \dot{\rho_d}+3H(\rho_d+p_d)+Q(\rho_d,\rho_m)=0,
\end{equation}
\begin{equation}
    \dot{\rho_m}+3H\rho_m-Q(\rho_d,\rho_m)=0.
\end{equation}
For the pressure of dark fluid we again use the equation of state in the form (\ref{6}). The function $Q$  describes the interaction between the components. For example, let $Q$ be
\begin{equation}\label{fQ}
    Q=\delta (\rho_c-\rho_d) \rho_d, \quad \delta=\mbox{const}
\end{equation}
For some $\delta$ and $\Delta \xi$ one can construct the models with the exit from the inflation and transition between rapid acceleration and deceleration. On Fig. 4 we give some examples of such models for which $w\approx 0$ after exit from the inflation. Some amount of matter appears due to the interaction between dark energy and matter if we start from the moment when $\rho_m=0$. Then parameter $w_d$ for dark energy asymptotically tends to zero. In some sense we have usual baryon matter and dark matter. One can in principle obtain required relation between density of baryon matter and dark energy (see left upper panel of Fig. 4).

\begin{figure}
    \centering
    \includegraphics[scale=0.30]{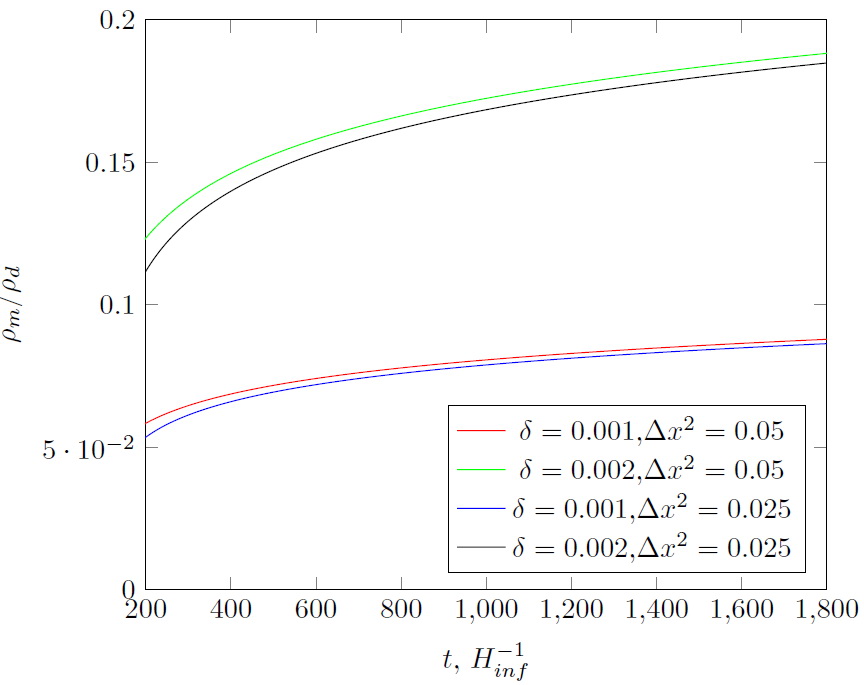}
    \includegraphics[scale=0.30]{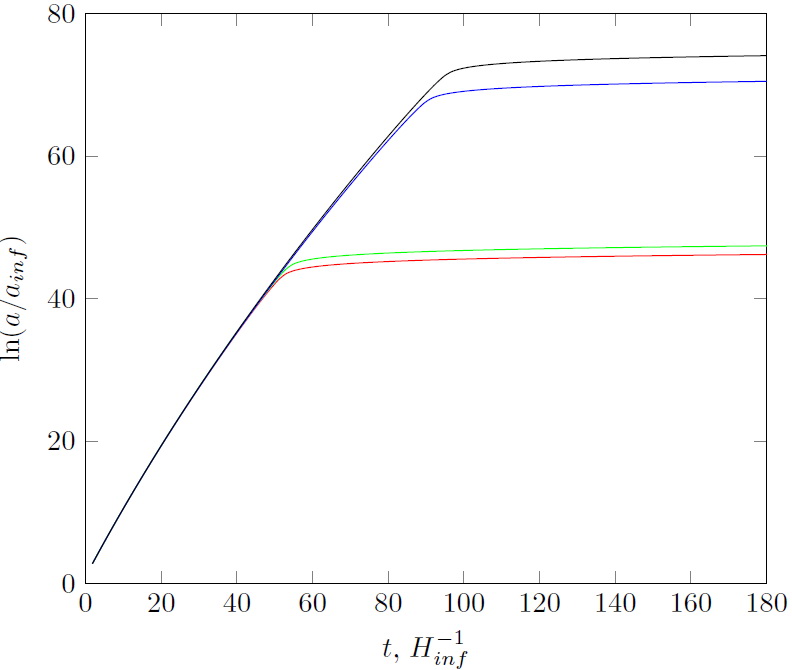}\\
    \includegraphics[scale=0.30]{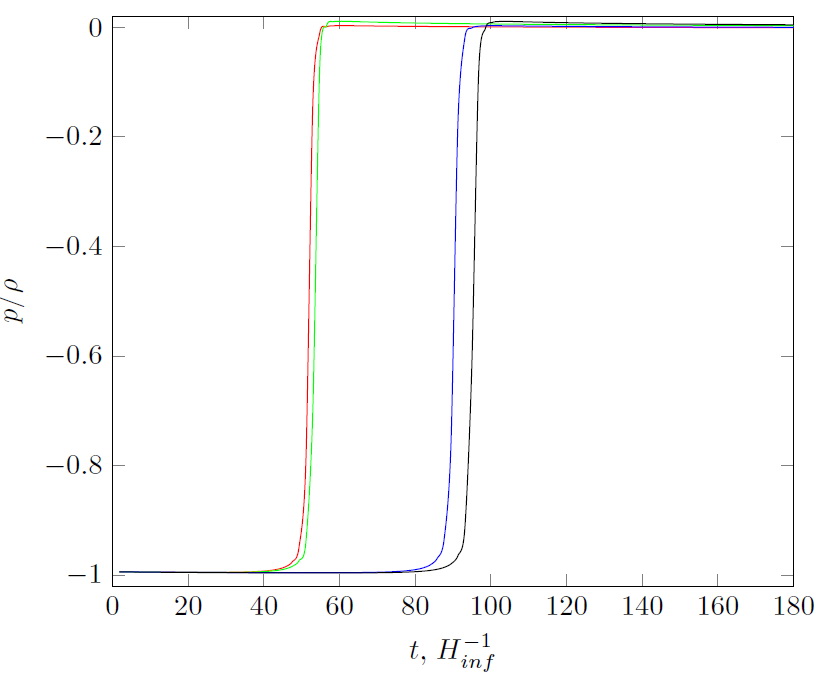}\\
    \caption{The relation between the matter density and dark energy density (upper left panel) for late times, number of e-foldings as function of time (upper right panel), parameter $\frac{p}{\rho_d+\rho_m}$ (down panel) in the model with interaction between matter and viscous fluid. Other parameters of the model are $\alpha=0.005$, $\beta=0.05$, $w_0=0.145$, $\xi_0=3.42$ and $\Delta\xi=0.88\xi_0$ for $\delta=0.001$ ($\Delta\xi=0.89\xi_0$ for $\delta=0.002$). We choose $\xi_{0}$ and $\Delta\xi$ so that asymptotic value of $w$ is $0$.}
    \label{fig:04}
\end{figure}

{One note that we consider unification in sense that one equation of state for dark energy can describe early inflation and transition to matter domination epoch. Effective EOS parameter tends to $0$ after inflation for some parameters of our model and therefore viscosity fluid can mimic dark matter. Interaction with baryon matter leads to creation of some amount of baryon matter. In result after inflation we have baryon matter and viscous fluid with $w=0$ which can be considered as dark matter.} 

Of course these examples are only illustration but probably in this way one can solve dark matter problem (for review of dark matter problem see \cite{Hooper}).

\section{Late-time acceleration in the model of fluid with viscosity}

Due to the viscosity the cosmological acceleration decreases. For the equation of state considered above energy density of van der Waals fluid asymptotically approaches $0$ at $t\rightarrow\infty$.
Hubble parameter also tends to zero. However, it is possible to construct solution with second phase of slow acceleration. For this one needs, for example, to change the function $f(\rho_d)$ in the equation of state as  
\begin{equation}\label{ff}
    f(\rho_d)=-\frac{\alpha \rho_{d}^{2}}{\rho_c}-{\delta_1}{\rho_c} \exp(-\delta_2\rho_{d}/\rho_c),
\end{equation}
where $\delta_1$ is small dimensionless positive value and $\delta_2$ is some constant. The value $\rho_{c}$ is used only for dimensional reasons in (\ref{ff}). At the beginning of the inflation this term is very small but it plays role at small energy densities. The possible physical motivation for this choice of course is not clear. But this is simplest way to modify EOS of viscous fluid for obtaining of late-time acceleration. 

One need to stress that the EOS (\ref{ff}) doesn't coincide with standard EOS for late time acceleration. For $\rho_{d}<<\rho_c$ we have that pressure of dark energy is 
$$
p_{d}\approx w_{0}\rho_{d} - \xi\rho_{d}/3 - C,\quad C=\delta_{1}\rho_{c}.
$$
The Eq.(\ref{eqln}) is rewritten in this case as
\begin{equation}
 \frac{d\ln a}{dx}=-\frac{(1-\beta x)}{x(a_{1}x^2+a_{2}x+a_{3}+d(x)+g(x))},
\end{equation}
$$
g(x)=-\frac{\delta_1}{x} (1-\beta x) \exp(-\delta_2 x).
$$
The main point is that for some small $x$ the expression in denominator has a zero at some $x=x_f$. Therefore if inflation begins from some density $x_{inf}$ for which $a_{1}x_{inf}^2+a_{2}x_{inf}+a_{3}+d(x_{inf})+g(x_{inf})>0$, the energy density asymptotically approaches $x_f$ at $t\rightarrow\infty$. The Universe expands according to de Sitter law. 
$$
a\sim \exp(\sqrt{\rho_f/3} t),\quad \rho_f\approx \delta_1 \frac{\rho_c}{c+d(0)}.
$$
We have two epochs of acceleration, one is the inflation (when parameter of equation-of-state is close to $-1$) and second one is de Sitter expansion (again the equation-of-state parameter is $\approx-1$). 

As on only illustrative examples we consider the cases when $\delta_1=10^{-6}$, $10^{-7}$ and $\delta_1=10^{-8}$ and include in our model the interaction between viscous fluid and matter in form (\ref{fQ})(see Fig. 5). At late times (for $\delta_1=10^{-8}$ after $\approx 1000H_{inf}^{-1}$) the EoS parameter approaches $-1$. The rate of the expansion is relatively slow (in comparison with the early inflation). 

\begin{figure}
    \centering
    \includegraphics[scale=0.3]{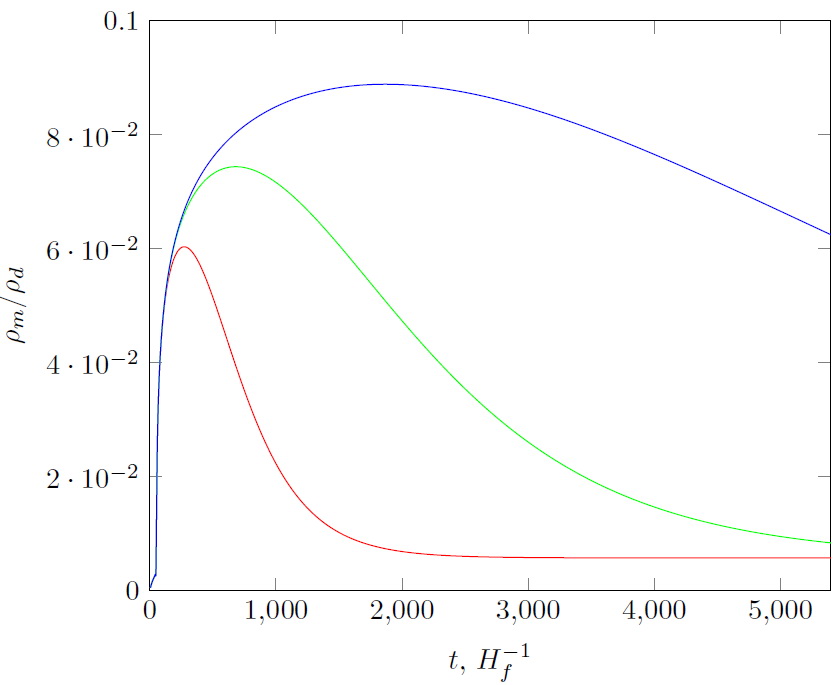}\includegraphics[scale=0.3]{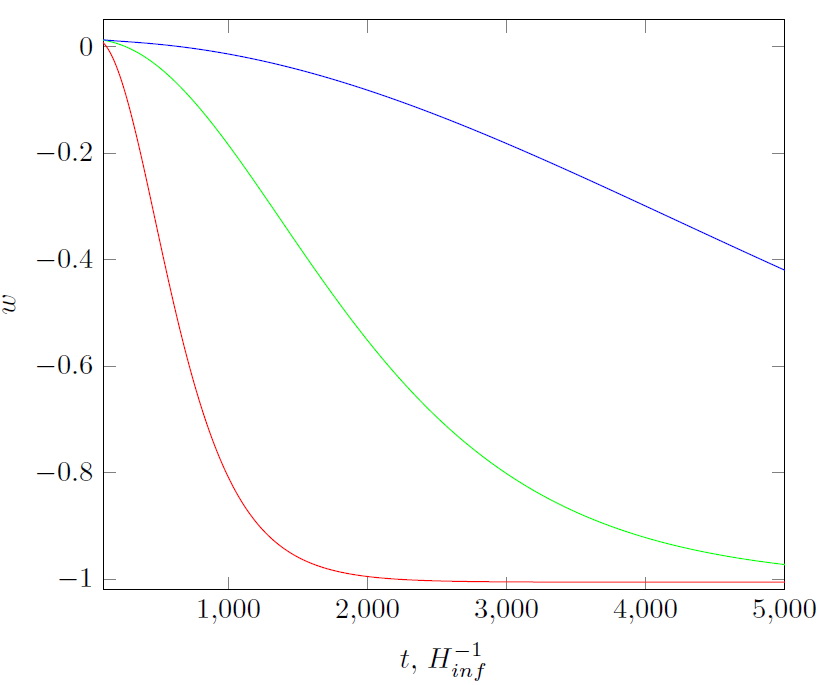}\\
    \includegraphics[scale=0.3]{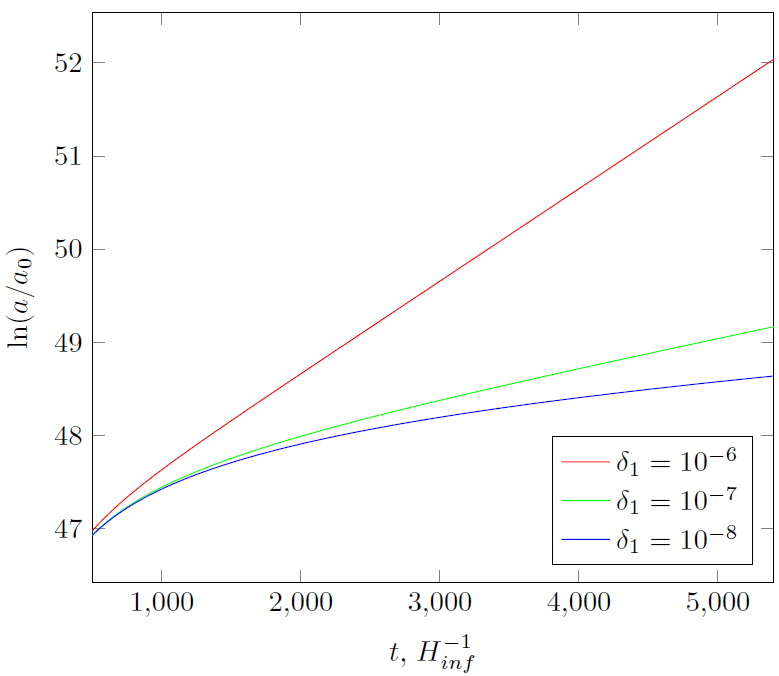}
    \caption{The dependence of e-foldings (upper left panel), parameter of state (upper right panel) from time after initial fast acceleration and dependence of relation $\rho_m/\rho_d$ from time for various values of $\delta_1$ and interaction between matter and viscous fluid. Other parameters of the model are $\alpha=0.005$, $\beta=0.05$, $w_0=0.145$, $\xi_0=3.42$ and $\Delta\xi=0.88\xi_0$, $\delta=0.001$, $\delta_2=\rho_c^{-1}.$}
    \label{fig:06}
\end{figure}

Of course the considered values of $\delta_{1}$ are non realistic.  The current Hubble parameter $H_{0}$ is many orders of magnitude smaller than $H_{inf}$. However, simple model (\ref{6}) with function $f$ (\ref{ff}) for $\rho_d<<\rho_c$ can describe observational data. One notes that after inflation $\rho_d<<\rho_c$ and therefore the equation of state with good accuracy can be written as (up to zero order in small parameter $\rho_d/\rho_c$)
\begin{equation}\label{model_real}
    p_d=w_{0}\rho_d - \zeta \rho/3 + p_f,\quad p_f=-\delta_1\rho_c.
\end{equation}
{{Here we designate $\zeta\approx \xi_0-\Delta \xi$. This is a value of viscosity after the phase transition}. Let's study cosmological evolution for this model. We compared our results with the following observational data:

(i) the dependence between magnitude and redshift for  Ia supernovae from the Supernova Cosmology Project,

(ii) the dependence of Hubble parameter from redshift, obtained from cosmic chronometry and baryon acoustic oscillation data.

Let's briefly look on these observational constraints in detail.

(i) The visual magnitude $\mu(z)$ for a supernova with redshift
$z=a_{0}/a-1$ is 
\be 
\mu(z)=\mu_{0}+5\log D(z)\, , 
\ee 
where
$D_{L}(z)$ is the luminosity distance:
\be
\label{DLSC} D_{L}(z)=\frac{c}{H_{0}}(1+z)\int_{0}^{z} h^{-1}(z)d
z, \quad h^{2}(z)=\rho(z)/\rho_{0}. \ee 
Here $c$ is speed of light
and $H_{0}$ is Hubble parameter at current moment of time. The best fit
for SNe is given in the framework of $\Lambda$CDM cosmology. For such a model (``standard cosmology''), one
obtains \be h(z)=(\Omega_{m}(1+z)^{3}+\Omega_{\Lambda})^{1/2}
\ee Here, $\Omega_{m}$ is the fraction of the total density
contributed by matter at present time, and $\Omega_{\Lambda}=1-\Omega_{m}$ is
the fraction contributed by the vacuum energy density.
The constant value $\mu_{0}$ depends on the chosen Hubble parameter:
$$
\mu_{0}=42.384-5\log h,\quad h=H_{0}/100 \mbox{km/s/Mpc}
$$
For the analysis of the SNe data one needs to calculate the parameter $\chi^{2}$, which is defined in the standard way
\begin{equation}
\chi^{2}_{SN}=\sum_{i}\frac{(\mu_{obs}(z_{i})-\mu_{th}(z_{i}))^{2}}{\sigma^{2}_{i}},
\end{equation}
where $\sigma_{i}$ is the corresponding $1\sigma$ error. We use data for 580 SNe Ia samples from \cite{Amman}. 

The parameter $\mu_{0}$ is independent of the data points and, therefore we can perform the minimization of $\chi^2$ with respect to $\mu_{0}$. One notes that
\begin{equation}\label{chi}
\chi^{2}_{SN}=A-2\mu_{0}B+\mu_{0}^{2}C,
\end{equation}
where
$$
A=\sum_{i}\frac{(\mu_{obs}(z_{i})-\mu_{th}(z_{i};\mu_{0}=0))^{2}}{\sigma^{2}_{i}},
$$
$$
B=\sum_{i}\frac{(\mu_{obs}(z_{i})-\mu_{th}(z_{i}))}{\sigma^{2}_{i}},\quad C=\sum_{i}\frac{1}{\sigma^{2}_{i}}.
$$
The $\chi$-square value (\ref{chi}) has a minimum for $\mu_{0}=B/C$ and this minimum is
$$
\bar{\chi}_{SN}^{2}=A-B^{2}/C.
$$
One can minimize $\bar{\chi}_{SN}^{2}$ instead of ${\chi}_{SN}^{2}$ and also compare optimal value of $\mu_0=B/C$ with the value for $H_{0}=73.24\pm
1.74$ km/s/Mpc from recent observations \cite{Riess2016}.  

$1\sigma$ and $2\sigma$ confidence level are defined by the following relations
$$
\Delta\chi^{2}=\chi^{2}-\chi^{2}_{min}<1.0 \mbox{ or } 4.0
$$ 
for the one-parametric model. For two-parametric model one need to change these values on $2.3$ and $6.17$ correspondingly \cite{Nesseris}.

(ii) There are various methods for measuring Hubble parameter as function of $z$. The largest amount of data was obtained using cosmic chronometric method. Hubble parameter depends on the differential age of the universe as a function of the redshift:
$$
dt=-\frac{1}{H}\frac{dz}{1+z}.
$$
Measurements of $dz/dt$ (and measurement of $H(z)$ as consequence) are possible due to data about absolute age for passively evolving galaxies, determined from fitting stellar population models. These measurements are given in papers \cite{Zhang}, \cite{Simon}, \cite{Moresco2}, \cite{Moresco3}, \cite{Stern}, \cite{Ratsimbazafy}.

There are also three correlated measurements of $H(z)$
from the radial BAO signal in the galaxy distribution \cite{Alam} and two values for high redshif ($z=2.34$ and 2.36) measured from the BAO signal in the Lyman-alpha forest distribution \cite{Delubac}, \cite{Font-Ribera}.

We use these 36 measurements of $H(z)$ compiled by \cite{Yu}. These data are listed in Table I.

\begin{table}
\label{Table1}
\begin{centering}
\begin{tabular}{|c|c|c|c|}
  \hline
  $z$ & $H_{obs}(z)$ & $\sigma_{H}$  \\
      & km s$^{-1}$ Mpc$^{-1}$        &  km s$^{-1}$ Mpc$^{-1}$     \\
  \hline
  0.070 & 69 & 19.6\\
  0.090 & 69 & 12 \\
  0.120 & 68.6 & 26.2 \\
  0.170 & 83 & 8  \\
  0.179 & 75 & 4\\
  0.199 & 75 & 5\\
  0.200 & 72.9 & 29.6\\
  0.270 & 77 & 14 \\
  0.280 & 88.8 & 36.6\\
  0.352 & 83 & 14\\
  0.38 & 81.9 & 1.9\\
  0.3802 & 83 & 13.5\\
  0.400 & 95 & 17 \\
  0.4004 & 77 & 10.2\\
  0.4247 & 87.1 & 11.2\\
  0.4497 & 92.8 & 12.9\\
  0.470 & 89 & 50\\
  0.4783 & 80.9 & 9\\
  0.480 & 97 & 62 \\
  0.510 & 90.8 & 1.9\\
  0.593 & 104 & 13\\
  0.610 & 97.8 & 2.1\\
  0.68 & 92 & 8\\
  0.781 & 105 & 12\\
  0.875 & 125 & 17\\
  0.880 & 90 & 40 \\
  0.900 & 117 & 23 \\
  1.037 & 154 & 20 \\
  1.300 & 168 & 17 \\
  1.363 & 160 & 33.6\\
  1.430 & 177 & 18 \\
  1.530 & 140 & 14 \\
  1.750 & 202 & 40 \\
  1.965 & 186.5 & 50.4\\
  2.34 & 223 & 7\\
  2.36 & 227 & 8\\
  \hline
\end{tabular}
\caption{Hubble parameter versus redshift data from \cite{Yu}.}
\end{centering}
\end{table}

The parameter $\chi^{2}_{H}$ is
\begin{equation}
\chi^{2}_{H}=\sum_{i}\frac{(H_{obs}(z_{i})-H_{th}(z_{i}))^{2}}{\sigma^{2}_{i}}.
\end{equation}
We also can perform marginalization over parameter $H_{0}$. The expansion of previous equation gives
$$
\chi^{2}_{H}=A_{1}-2B_{1}H_{0}+H_{0}^{2}C_{1},
$$
$$
A_{1}=\sum_{i}\frac{H_{obs}(z_{i})^{2}}{\sigma^{2}_{i}},\quad B_{1}=\sum_{i}\frac{h(z_{i})H_{obs}(z_{i})}{\sigma^{2}_{i}},\quad
$$
$$
C_{1}=\sum_{i}\frac{h(z_{i})^2}{\sigma^{2}_{i}}.
$$
For $H_{0}=B_{1}/C_{1}$ parameter $\chi^{2}_{H}$ is minimal.
$$
\bar{\chi}_{H}^{2}=A_{1}-B_{1}^{2}/C_{1}.
$$
As in the case of the SNe data, one can find minimum of $\bar{\chi}_{H}^{2}$ instead of ${\chi}_{H}^{2}$.

{We analysed model (\ref{model_real}) for fixed $w_0$ and some values of $\zeta$ varying remaining two parameters namely $p_f$ and $\Omega_{d}$. From our analysis it follows that for some small values of $\zeta$ there is intersection between allowed areas of free parameters for SNe data and data about $H(z)$.} 

{We calculated $68.3$\% and $95$\% allowed areas on plane $p_f-\Omega_{de}$ (see Fig. \ref{fig:071}). One notes also interesting moment. As well known, the tension between SNe data and $H(z)$ dependence exists for $\Lambda$CDM model. The optimal values of $H_{0}$ for two datasets differ significantly. In the considered model this discrepancy also takes place but not so dramatically. For SNe data $H_{0}\sim 70$ km/s/Mpc gives for $\chi^2_{SN}$ value which should be compared with $\Lambda$CDM model. The corresponding value for $\chi^2_{H}$ lies in the appropriate limits for this $H_{0}$. The key point of model (\ref{model_real}) is that for SNe data we have very large interval for $\Omega_d$ and good agreement with data is possible for various $H_0$.}   

\begin{figure}
    \centering
    A) $\zeta=0.009$\\
    \includegraphics[scale=0.3]{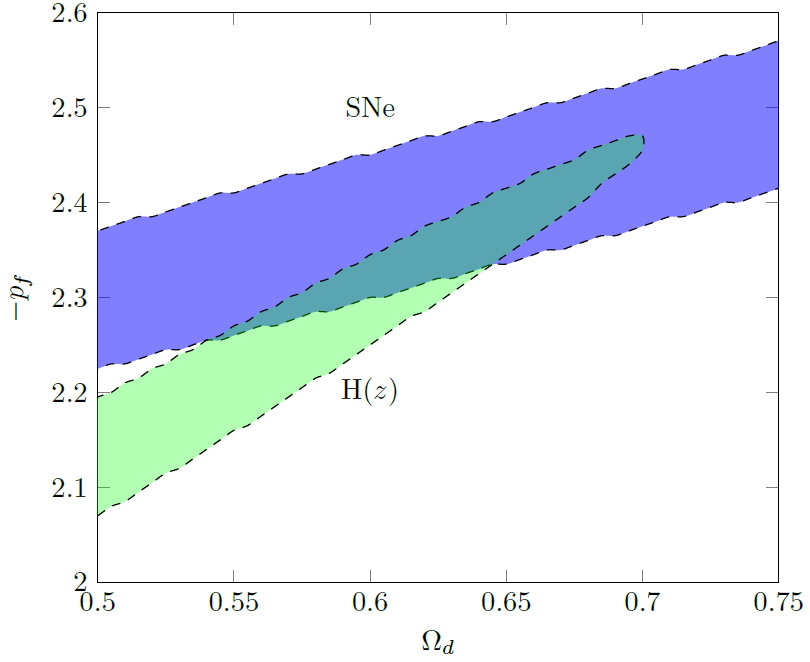}
    \includegraphics[scale=0.3]{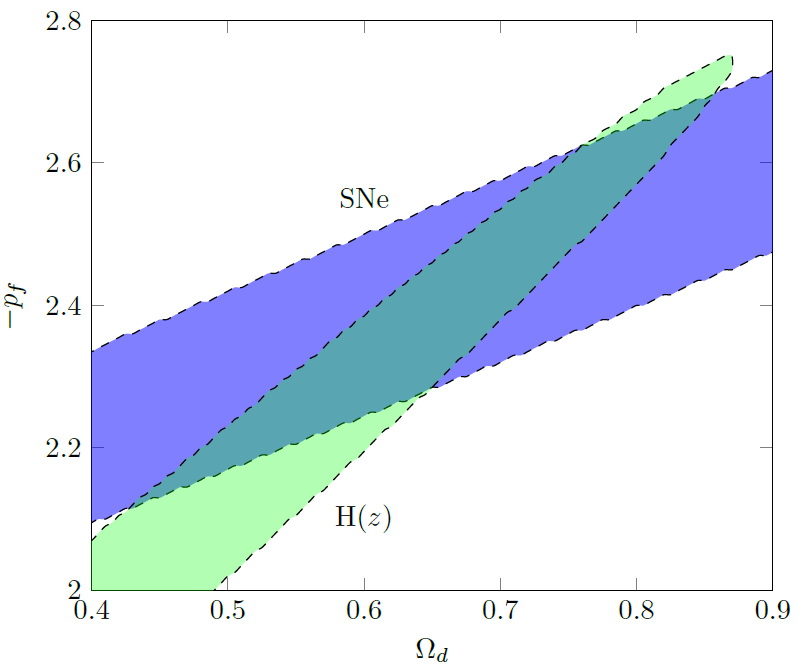}\\
    B) $\zeta=0.09$\\
    \includegraphics[scale=0.3]{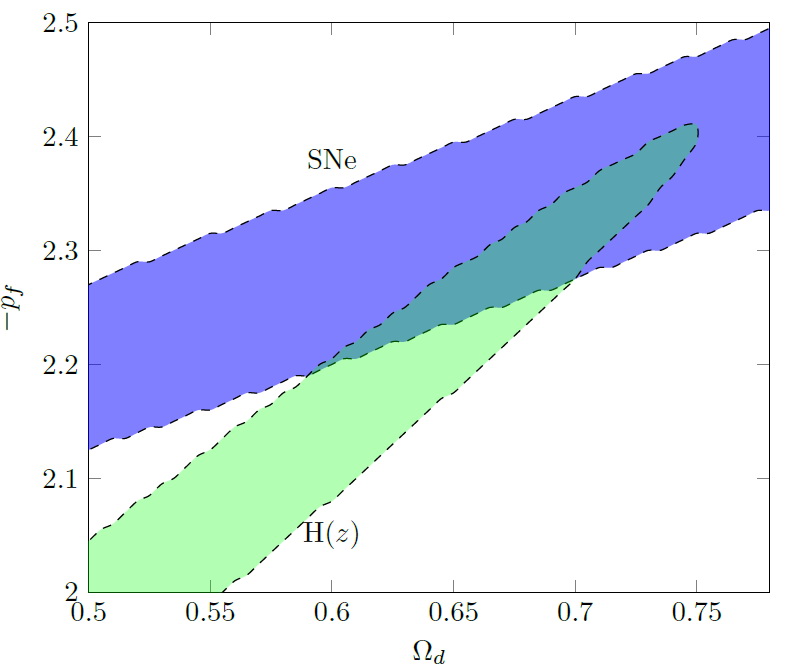}
    \includegraphics[scale=0.3]{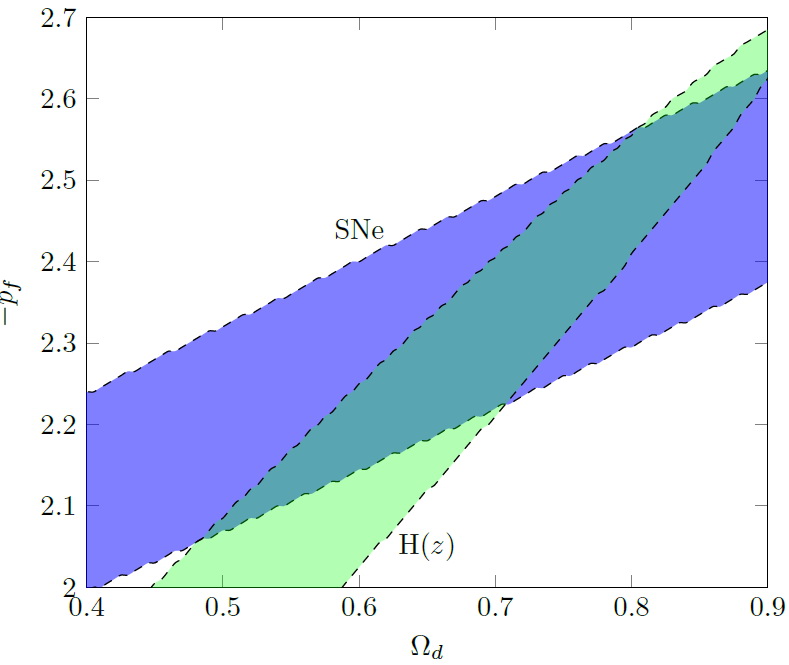}
    \caption{$1\sigma$ (left panel) and $2\sigma$ (right panel) allowed areas on plane $p_f-\Omega_{d}$ for model (\ref{model_real}). The values for $p_f$ are given in the units of $H_0^{2}$.}
    \label{fig:071}
\end{figure}

\section{Comparison with Planck observations}

\begin{figure}
    \centering
    \includegraphics[scale=0.3]{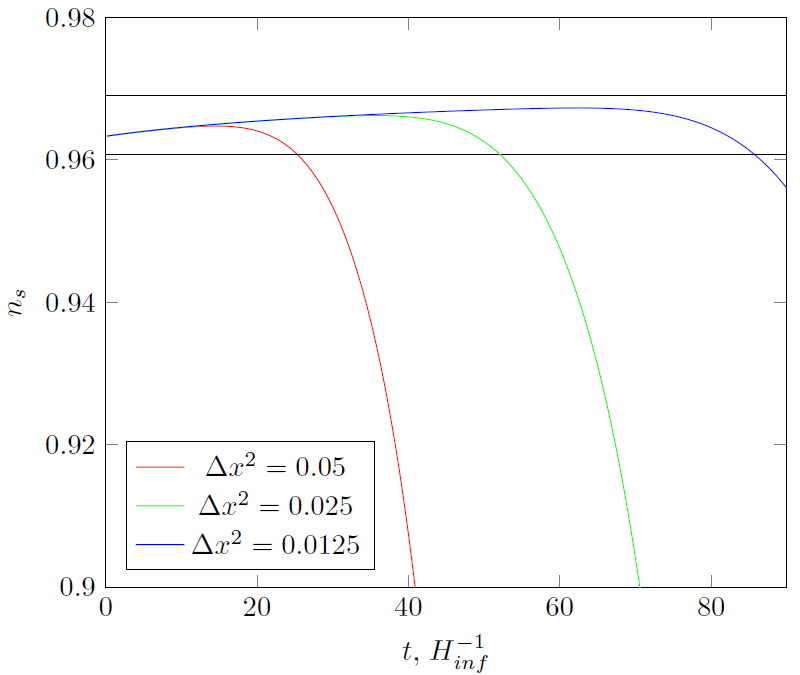}
    \includegraphics[scale=0.3]{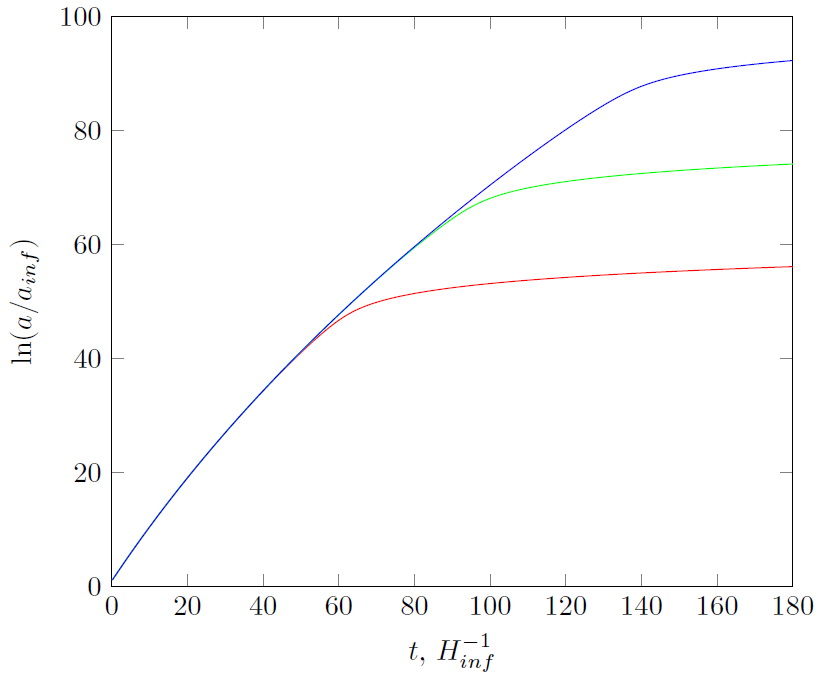}
    \caption{The dependence of the spectral index $n_s$ from time (left panel) for various values of $\Delta x^2$ in the initial period of fast acceleration with constraints from Planck observations. Other parameters of model are $\alpha=0.005$, $\beta=0.05$, $w_0=0.145$, $\xi_0=3.42$ and $\Delta\xi=0.2\xi_0$, $\rho_f=0$. On right panel the dependence of e-foldings from time is given. For concordance with bounds on spectral index the fluctuations of matter density should appear during approximately half of period of fast acceleration.}
    \label{fig:07}
\end{figure}

We will now compare the predictions of our inflationary model with the latest
Planck satellite observational data. In particular, we will calculate the
parameters of inflation, and consider how the spectral index match with the values obtained from the astronomical
data analysis.

We consider the evolution of the scale factor when viscosity is $\xi\sim H$. The scale factor is then given by the expression (\ref{eqln}).  We
intend to see under what conditions the inflationary model complies with the Planck data. Of course one can neglect additional term in (\ref{ff}) because $\delta_1$ is approximately equal to relation between energy density in the epoch of inflation and its current value i.e. $\delta_{1}$ is $\sim 10^{-100}$.

Let us calculate the ``slow-roll'' parameters $\varepsilon$ and $\eta$ for considered model. As is known
\begin{equation}
\varepsilon=-\frac{\dot{H}}{H^2}=\frac{3}{2}\left(1+\frac{p_d}{\rho_d}\right). \quad
 \label{25}
\end{equation}
 For the model with EoS (\ref{4}) we have
 $$
\varepsilon=\frac{3}{2}\left(1+\frac{w_{0}}{1-\beta x}-\alpha x-\frac{\xi_0}{3}+\frac{\Delta\xi}{3}\exp(-(x-x_f)^2/{\Delta x}^2)\right)
 $$
Parameter $\eta$ according to the definition is
\begin{equation}
    \eta= \varepsilon - \frac{1}{2 \varepsilon H} \dot{\varepsilon}.
\end{equation}
After simple calculations one can obtain
\begin{equation}
    \eta = \varepsilon + \frac{d\varepsilon}{d\rho_d} \rho_d=
\end{equation}
$$
=\frac{3}{2}\left(1+\frac{w_0}{(1-\beta x)^2}-2\alpha x -\frac{\xi_0}{3}+\frac{\Delta\xi}{3}(1-2x(x-x_f){\Delta x}^{-2}\exp(-(x-x_f)^2/{\Delta x}^2)\right).
$$
For slow-roll inflation parameters $\epsilon$ and $\eta$  should be small. In this case spectral index of scalar perturbations is equal
\begin{equation}
    n_s = 1 - 6\varepsilon + 2\eta
\end{equation}
or
$$
n_s = 1 - 4\varepsilon + 2 \varepsilon' x = 
$$
$$
=1 - 3\left(2-\frac{2\xi_0}{3}+\frac{w_0(2-3\beta x)}{(1-\beta x)^2}-\alpha x +\frac{2}{3}\Delta \xi \exp(-(x-x_f)^2/{\Delta x}^2)\left[1-\frac{x(x-x_f)}{{\Delta x}^2}\right]\right)
$$
According to observations from Planck, the spectral index lies within narrow limits:
$$
n_s = 0.9649 \pm 0.0042.
$$
We assume that inflation begins when $x>>x_{f}$ and the term with $\Delta \xi$ is negligible. We also choose for simplicity values of parameters $\xi_0$, $w_0$, $\alpha$, $\beta$ for which $4a_{1}a_{3}-a_{2}^2=0$. The dependence of spectral index from time is depicted on Fig. 7. From this dependence it follows that one needs to assume that matter density fluctuations should form at least during approximately half of time when fast acceleration ends. In this case the scalar spectral index of perturbations lies within the limits of Planck data.
 
\section{Conclusions}

We proposed the unified description of the early acceleration of the universe (i.e. cosmological inflation)
and the current epoch of ``dark energy'' domination. The inflation can be described by cosmic fluid with viscous van der Waals EoS. 

Viscosity leads to slow-roll inflation with acceptable spectral index and another parameters of inflation constrained by Planck observations. Assuming the ``phase transition'' for which viscosity sharply decreases one can describe the exit from inflation.  

One can also construct the models in which dark fluid mimics dark matter with the effective EoS parameter close to $w\approx 0$. In some sense we mimic the dark matter. Interaction between viscous fluid and matter allows to construct the models with the exit from the inflation and transition between rapid acceleration and deceleration. One can, in principle, obtain the required relation between density of baryon matter and dark energy.

One can add simple term $\sim -p_f \exp(-\rho/\rho_c)$ into EoS the contribution of which is very small for the inflation but crucial for late acceleration. We showed that such model is well fitted with observational bounds from supernovae cosmology and dependence of Hubble parameter from redshift obtaining from cosmic chronometry and baryon acoustic oscillations data.

\medskip

\noindent {\bf Acknowledgements}. This work was supported by Ministry of Education and Science (Russia), project 075-02-2021-1748 (AVA, AST) and MINECO (Spain), project PID 2019-104397GB-I00 (SDO).

\end{document}